\documentclass [a4paper,fleqn, 12pt]{article}
\usepackage{graphicx}
\usepackage[small]{subfigure,epsfig}

\usepackage {amsmath} \usepackage{amssymb} \usepackage{cite}
\newcommand{\cn}

\begin{document}

%\begin{frontmatter}

\title
{From Laurent Series to Exact Meromorphic Solutions: the Kawahara equation}

\author
{Maria V. Demina, \and Nikolay A. Kudryashov}

\date{Department of Applied Mathematics, National Research Nuclear University
MEPHI, 31 Kashirskoe Shosse,
115409 Moscow, Russian Federation}

%\author

%\corauthref{cor}},
%\corauth[cor]{Corresponding author.} \ead{nakudr@gmail.com}

%\address

\maketitle

\begin{abstract}

Nonlinear waves are studied in a mixture of liquid and gas bubbles. Influence of viscosity and heat transfer is taken into consideration on propagation of the pressure waves. Nonlinear evolution equations of the second and the third order for describing nonlinear waves in gas-liquid mixtures are derived. Exact solutions  of these nonlinear evolution equations are found. Properties of nonlinear waves in a liquid with gas bubbles are discussed.

\end{abstract}

%\begin{keyword}

% Point vortices, Special polynomials, Adler -- Moser polynomials

%\PACS 02.30.Hq - Ordinary differential equations

%\end{keyword}

%\end{frontmatter}

\section{Introduction}

In this article we study meromorphic traveling wave solutions of the
following partial differential equations
\begin{equation}
\label{KW_4} u_t+\alpha u^nu_x+\beta u_{xxx}- \delta
u_{xxxxx}=0,\quad \alpha\neq 0,\quad \delta\neq0,
\end{equation}
where $n=1$, $n=2$, and $n=4$. In the case $n=1$ equation
\eqref{KW_4} is the famous Kawahara equation, arising in several
physical applications. For example, in the theory  of
magneto--acoustic waves in plasma \cite{Kawahara01} and in the
theory of long waves in shallow liquid under ice cover
\cite{Marchenko01}. In the case $n=2$ and $n=4$ equation
\eqref{KW_4} may be regarded as modifications of the Kawahara
equation. Let us call these equations the modified Kawahara
equations. Several families of exact solutions to the Kawahara
equation and the modified Kawahara equation \eqref{KW_4} with $n=2$
are given in \cite{Kano01, Kudryashov10, Parkes01, Parkes02,
Bagderina01, Biswas01, Natali01}.

The problem of constructing exact solutions for nonlinear
differential equations is intensively studied in recent years
\cite{Kudr91, Kudryashov04, Kudr92, Malfliet01, Parkes01, Parkes02,
Kudr05b, Kudryashov05, Kudryashov06, Kudr08a}. A lot of methods and
new families of exact solutions appear \cite{Kudryashov01,
Kudryashov02, Kudryashov03}. However, the question of classification
for exact solutions is addressed very seldom (see \cite{Conte01,
Hone01, Vernov02, Eremenko01, Demina01}). In this article we
describe a method, which can be used to classify and build in
explicit form meromorphic solutions of autonomous nonlinear ordinary
differential equations.

This article is organized as follows. In section 2 we give explicit
expressions for meromorphic solutions and describe a process of
their construction. In sections 3, 4, and 5 we find all meromorphic
solutions of third order ordinary differential equations satisfied
by traveling wave solutions of the Kawahara equation and the
modified Kawahara equations.

\section{Method applied} \label{Method applied}

Equation \eqref{KW_4} admits the traveling wave reduction
$u(x,t)=w(z)$, $z=x-C_0t$ with $w(z)$ satisfying the equation
\begin{equation}
\label{ODE4_KW_4}\delta w_{zzzz}-\beta w_{zz}- \frac{\alpha }{n+1}
w^{n+1}+C_0w+C_1=0,
\end{equation}
where $C_1$ is an integration constant. Multiplying equation
\eqref{ODE4_KW_4} by $w_z$ and integrating the result, we obtain the
equation
\begin{equation}
\label{ODE3}\delta
\left(w_{zzz}w_z-\frac12w_{zz}^2\right)-\frac{\beta}{2}
w_{z}^2-\frac{\alpha
}{(n+1)(n+2)}w^{n+2}+\frac{C_0}{2}w^2+C_1w+C_2=0,
\end{equation}
where again $C_2$ is an integration constant. Without loss of
generality, the parameters $\alpha$, $\delta$ in  \eqref{KW_4}  may
be taken arbitrary. Let us set
\begin{equation}
\begin{gathered}
\label{Parameters} n=1:\quad \delta=1,\quad \alpha=6;\hfill \\
n=2:\quad \delta=1,\quad \alpha=360;\hfill \\
n=4:\quad \delta=1,\quad \alpha=120.\hfill
\end{gathered}
\end{equation}
The aim of this article is to obtain all the families of meromorphic
solutions for equation \eqref{ODE3} with $n=1$, $n=2$, $n=4$. We
shall use an approach suggested in \cite{Demina01}. Consider an
autonomous nonlinear ordinary differential equation
\begin{equation}
\label{EQN} E[w(z)]=0,
\end{equation}
where $E[w(z)]$ is a polynomial in $w(z)$ and its derivatives such
that substituting $w(z)=\lambda W(z)$ into equation \eqref{EQN}
yields an expression with only one term of the highest degree in
respect of~$\lambda$. For every solution $w(z)$ of equation
\eqref{EQN} there exists a family of solutions $w(z-z_0)$. We shall
omit arbitrary constant~$z_0$. Suppose that equation \eqref{EQN}
possesses only one asymptotic expansion corresponding to the Laurent
series in a neighborhood of the pole $z=0$
\begin{equation}
\begin{gathered}
\label{Laurent_expantionN1}
w^{}(z)=\sum_{k=1}^{p}\frac{c_{-k}^{}}{z^k}+\sum_{k=0}^{\infty}c_k^{}z^k,\quad
0<|z|<\varepsilon,
\end{gathered}
\end{equation}
Here $p>0$ is an order of the pole $z=0$. For example, this is the
case of equation \eqref{ODE3} with $n=1$. Exact meromorphic
solutions of equation \eqref{EQN} satisfied by one formal Laurent
expansion \eqref{Laurent_expantionN1} are given in theorem 1 \cite{Demina01}.

\textbf{Theorem 1.}
 All meromorphic solutions of equation \eqref{EQN} with
only one asymptotic expansion \eqref{Laurent_expantionN1}
corresponding to the Laurent series in a neighborhood of the pole
$z=0$ are of the form:
\newline 1) elliptic solutions with the periods $2\omega_1$,
$2\omega_2$ and one pole of order $p$ inside a parallelogram of
periods
\begin{equation}
\begin{gathered}
\label{Ex_Sol_Elliptic2} w(z)=\left\{\sum_{k=2}^{p}\frac{(-1)^k
c_{-k}}{(k-1)!}\frac{d^{k-2}}{dz^{k-2}}\right\}\wp(z;\omega_1,\omega_2)+
h_0,
\end{gathered}
\end{equation}
Necessary condition for elliptic solutions to exist is $c_{-1}=0$.
\newline 2) simply periodic solutions with the period $T$ and one
pole of order $p$ inside a stripe of periods built on $T$
\begin{equation}
\begin{gathered}
\label{Ex_Sol_Expp} w(z)=
\frac{\pi}{T}\left\{\sum_{k=1}^{p}\frac{(-1)^{k-1}
c_{-k}}{(k-1)!}\frac{d^{k-1}}{dz^{k-1}}\right\}\cot \left(\frac{\pi
z}{T}\right)+ h_0.
\end{gathered}
\end{equation}
\newline 3) rational solutions
\begin{equation}
\begin{gathered}
\label{Rat_sol_N=1}
w(z)=\sum_{k=1}^{p}\frac{c_{-k}}{z^k}+\sum_{k=0}^{m}c_kz^k,\quad
m\geq 0,\hfill \\
\\
w(z)=\sum_{k=0}^{m}h_kz^k,\quad m\geq 0. \hfill
\end{gathered}
\end{equation}
In 1), 2), and 3) $h_k$, $0\leq k \leq m$ are constants and the
Weierstrass elliptic function $\wp$ satisfies the equation
\begin{equation}
\begin{gathered}
\label{Wier} (\wp_z)^2=4\wp^3-g_2\wp-g_3.
\end{gathered}
\end{equation}

Now suppose that equation \eqref{EQN} possesses $N$ different
asymptotic expansions corresponding to the Laurent series in a
neighborhood of the pole $z=0$
\begin{equation}
\begin{gathered}
\label{Laurent_expantionN}
w^{(i)}(z)=\sum_{k=1}^{p_i}\frac{c_{-k}^{(i)}}{z^k}+\sum_{k=0}^{\infty}c_k^{(i)}z^k,\quad
0<|z|<\varepsilon_i,\quad i=1, \ldots, N.
\end{gathered}
\end{equation}
In this expressions $p_i>0$ is an order of the pole $z=0$. For
instance, equation \eqref{ODE3} with $n=2$ admits two different
formal Laurent series and equation \eqref{ODE3} with $n=4$ admits
four different formal Laurent series. Let us call any pole $z=b$ of
meromorphic solution $w(z)$ with the Laurent expansion
$w^{(i)}(z-b)$ (see \eqref{Laurent_expantionN}) as a pole of type
$i$. Meromorphic solutions of equation \eqref{EQN} are classified in
theorem 1 \cite{Demina01}. Again we omit arbitrary constant
$z_0$.

\textbf{Theorem 2.}
 All meromorphic solutions of equation \eqref{EQN} with
$N$ different asymptotic expansions \eqref{Laurent_expantionN}
corresponding to the Laurent series in a neighborhood of the pole
$z=0$ are of the form:
\newline 1) elliptic solutions with the periods $2\omega_1$,
$2\omega_2$ and $|I|+1$ poles $\{a_i\}$ of orders $\{\text{ord}$
 $a_i= p_i\}$, $i\in I\cup$ $\{i_0\}$ inside a parallelogram of
periods
\begin{equation}
\begin{gathered}
\label{Ex_Sol_EllipticN} w(z)=\left\{\sum_{i\in\,
I}^{}\sum_{k=2}^{p_i}\frac{(-1)^k
c_{-k}^{(i)}}{(k-1)!}\frac{d^{k-2}}{dz^{k-2}}\right\}\left(\frac14\left[
\frac{\wp_z(z)+B_i}{\wp(z)-A_i}\right]^2-\wp(z)\right)\\
\\
+\sum_{i\in \,
I}^{}\frac{c_{-1}^{(i)}(\wp_z(z)+B_i)}{2\,(\wp(z)-A_i)}+\left\{\sum_{k=2}^{p_{i_0}}\frac{(-1)^k
c_{-k}^{(i_0)}}{(k-1)!}\frac{d^{k-2}}{dz^{k-2}}\right\}\wp(z)+ h_0,
\end{gathered}
\end{equation}
where $\wp(z)\stackrel{def}{=}\wp(z;\omega_1,\omega_2)$,
$a_{i_0}=0$, $A_i\stackrel{def}{=}\wp(a_i)$,
$B_i\stackrel{def}{=}\wp_z(a_i)$, $B_i^2=4A_i^3-g_2A_i-g_3$, $i\in
I$. Necessary condition for elliptic solutions
\eqref{Ex_Sol_EllipticN} to exist is
\begin{equation}\label{Condition_ellipticN}
\sum_{i\in\,I}^{}c_{-1}^{(i)}+c_{-1}^{(i_0)}=0.
\end{equation}
\newline 2) simply periodic solutions with the period $T$ and
$|I|+1$ poles $\{a_i\}$ of orders $\{\text{ord}$
 $a_i= p_i\}$, $i\in I\cup$ $\{i_0\}$
inside a stripe of periods build on $T$
\begin{equation}
\begin{gathered}
\label{Ex_Sol_ExppN} w(z)=\sqrt{L}\left\{\sum_{i\in\,
I}^{}\sum_{k=1}^{p_i}\frac{(-1)^{k-1}
c_{-k}^{(i)}}{(k-1)!}\frac{d^{k-1}}{dz^{k-1}}\right\}\frac{A_i\cot
\left(\sqrt{L}z\right)+\sqrt{L}}{A_i-\sqrt{L}\cot \left(\sqrt{L}z\right)}\\
\\
+\sqrt{L}\left\{\sum_{k=1}^{p_{i_0}}\frac{(-1)^{k-1}
c_{-k}^{(i_0)}}{(k-1)!}\frac{d^{k-1}}{dz^{k-1}}\right\}\cot
\left(\sqrt{L}z\right)+ h_0,
\end{gathered}
\end{equation}
where $a_{i_0}=0$, $L\stackrel{def}{=}\pi^2/T^2$,
$A_i\stackrel{def}{=}\sqrt{L}\cot(\sqrt{L}a_i)$.
\newline 3) rational solutions
\begin{equation}
\begin{gathered}
\label{Rat_sol_N}
w(z)=\sum_{k=1}^{p_{i_0}}\frac{c_{-k}^{(i_0)}}{z^k} +\sum_{i\in\,
I}^{}\sum_{k=1}^{p_i}\frac{c_{-k}^{(i)}}{(z-a_i)^k}+\sum_{k=0}^{m}h_kz^k,\quad
m\geq 0\hfill \\
\\
w(z)=\sum_{k=0}^{m}h_kz^k,\quad m\geq 0. \hfill
\end{gathered}
\end{equation}
In 1), 2) and 3) $h_k$, $0\leq k \leq m$ are constants and
$I=\varnothing$ or $I$ $\subseteq$ $\{1,2,\ldots, N\}$
$\setminus\{i_0\}$, $1\leq i_0 \leq N$.

The proof of these theorems is based on the Mittag--Leffler's
expansions for meromorphic functions (see \cite{Demina01}).
Necessary condition \eqref{Condition_ellipticN} for existence of
elliptic solutions follows from the theorem for total sum of the
residues at poles inside a parallelogram of periods of an elliptic
function. Invariants $g_2$, $g_3$ such that $g_2^3-27g_3^2\neq0$
uniquely determine the elliptic function $\wp(z)$. In addition we
have the following correlations
\begin{equation}
\begin{gathered}
\label{Invariants} g_3={\sum_{(n,\,m)\neq(0,\,0)}}
\frac{60}{(2n\omega_1-2m\omega_2)^4},\,\quad
g_2={\sum_{(n,\,m)\neq(0,\,0)}}
\frac{140}{(2n\omega_1-2m\omega_2)^6}.
\end{gathered}
\end{equation}
In the case $g_2^3-27g_3^2=0$ the elliptic function $\wp(z)$
degenerates and consequently elliptic solutions
\eqref{Ex_Sol_EllipticN} degenerate. Note that for fixed values of
parameters in equation \eqref{EQN}, if any, there may exist only one
meromorphic solution (rational, simply periodic or elliptic) with a
pole at $z=0$ of type $i$, $1\leq i \leq N$.

Thus we see that the problem of constructing exact meromorphic
solutions of equation \eqref{EQN} in explicit form reduces to the
question of finding coefficients for solutions given in theorems
1 and 2. The basic idea is to expand these exact
solutions in a neighborhood of their poles and to compare
coefficients of these expansions with coefficients of series
\eqref{Laurent_expantionN}.

At \textit{the first step} for solutions of equation \eqref{EQN} one
constructs asymptotic expansions corresponding to Laurent series in
a neighborhood of the pole $z=0$.

At \textit{the second step} one selects an expression for a
meromorphic solution $w(z)$(see theorem 2). If the
meromorphic solution $w(z)$ possesses poles of types $i\in J$,
$J\subseteq$ $\{1,2,\ldots, N\}$, then one may take any $i\in J$ in
capacity of $i_{0}$ and suppose that the point $z=0$ is the pole of
type $i_0$ for $w(z)$.

At \textit{the third step} one expands the meromorphic solution
$w(z)$ in a neighborhood of the poles $\{a_i\}$, $i\in I\cup$
$\{i_0\}$, $a_{i_0}=0$. For elliptic solutions with the poles
$\{a_i\}$ of orders $\{\text{ord}$
 $a_i= p_i\}$, $i\in I\cup$ $\{i_0\}$  inside a
parallelogram of periods one takes the expression
\begin{equation}
\begin{gathered}
\label{Ex_Sol_EllipticN_proof} w(z)=\sum_{i\in
I\cup\{i_0\}}^{}c_{-1}^{(i)}\zeta(z-a_i)+\left\{\sum_{i\in
I\cup\{i_0\}}^{} \sum_{k=2}^{p_i}\frac{(-1)^k
c_{-k}^{(i)}}{(k-1)!}\frac{d^{k-2}}{dz^{k-2}}\right\}\wp(z-a_i)
+\tilde{h}_0,
\end{gathered}
\end{equation}
where $\tilde{h}_0$ is a constant, $a_{i_0}=0$, and $\zeta(z)$ is
the Weierstrass $\zeta$--function, finds the Laurent series for
$w(z)$ given by \eqref{Ex_Sol_EllipticN_proof} around its poles
$z=0$, $z=a_i$, $i\in I$. And then introduces notation
$A_i\stackrel{def}{=}\wp(a_i)$, $B_i\stackrel{def}{=}\wp_z(a_i)$,
$i\in I$, $h_0\stackrel{def}{=}\tilde{h}_0- \sum_{i\in\,
I}^{}c_{-1}^{(i)} \zeta(a_i)-\sum_{i\in \,I}^{}c_{-2}^{(i)}
\wp(a_i)$. Expression \eqref{Ex_Sol_EllipticN_proof} can be
rewritten as \eqref{Ex_Sol_EllipticN} with the help of addition
formulae for the functions $\wp$ and $\zeta$. For simply periodic
solutions with the poles $\{a_i\}$ of orders $\{\text{ord}$
 $a_i= p_i\}$, $i\in I\cup$ $\{i_0\}$  inside a
stripe of periods one takes the expression
\begin{equation}
\begin{gathered}
\label{Ex_Sol_ExppN_proof} w(z)= \frac{\pi}{T}\left\{\sum_{i\in
I\cup\{i_0\}}^{}\sum_{k=1}^{p_i}\frac{(-1)^{k-1}
c_{-k}^{(i)}}{(k-1)!}\frac{d^{k-1}}{dz^{k-1}}\right\}\cot
\left(\frac{\pi (z-a_i}{T}\right)+ h_0,
\end{gathered}
\end{equation}
where $a_{i_0}=0$, expands this function in a neighborhood of its
poles $z=0$, $z=a_i$, $i\in I$, and introduces notation
$L\stackrel{def}{=}\pi^2/T^2$,
$A_i\stackrel{def}{=}\sqrt{L}\cot(\sqrt{L}a_i)$. Expression
\eqref{Ex_Sol_ExppN_proof} can be rewritten as \eqref{Ex_Sol_ExppN}.
Note that it may be taken any value for square root of $L$, for
example, with sign $+$.

At \textit{the fourth step} one compares coefficients of the series
found at the first and the third steps and solves an algebraic
system for the parameters of exact meromorphic solution. In addition
correlations for the parameters of equation \eqref{EQN} may arise.
Optimal number of equations in the algebraic system equals to the
number of parameters of the meromorphic solution and equation
\eqref{EQN} plus $1$. If this system is inconsistent, then equation
\eqref{EQN} does not possess meromorphic solutions with supposed
expression. The parameters of exact meromorphic solution
\eqref{Ex_Sol_EllipticN} that one needs to calculate are $g_2$,
$g_3$, $h_0$, $A_i$, $B_i$ $i\in I$ in the case of elliptic
solutions, $h_0$, $L$, $A_i$ $i\in I$ in the case of simply periodic
solutions, and $a_i$, $i\in I$, $\{h_k\}$ in the case of rational
solutions. In order to find the highest exponent $m$ (see
\eqref{Rat_sol_N}) one may construct the Laurent expansion in a
neighborhood of infinity for $w(z)$.

At \textit{the fifth step} one verifies the meromorphic solution
obtained at the previous step, substituting this solution and
correlations on the parameters of equation \eqref{EQN} into the
latter.

In order to construct all the meromorphic solutions for equation
\eqref{EQN} one should consider all the variants for possible types
of poles possesses by $w(z)$, i.e. all subsets of $\{1,2,\ldots,
N\}$. Finally, we would like to mention that our approach may be
applied to build exact meromorphic solutions of autonomous nonlinear
ordinary differential equations with arbitrary constants in Laurent
series in a neighborhood of poles. However these equations may
possess meromorphic solutions of more complicated structure.

\section{Exact solutions of the Kawahara equation}

In this section we find all cases when the following third order
ordinary differential equation
\begin{equation}
\label{ODE3_KW_4}w_{zzz}w_z-\frac12w_{zz}^2-\frac{\beta}{2}
w_{z}^2-w^3+\frac{C_0}{2}w^2+C_1w+C_2=0,
\end{equation}
possesses meromorphic solutions and these solutions themselves.
Equation \eqref{ODE3_KW_4} arises as traveling wave reduction of the
Kawahara equation (equation \eqref{KW_4} with $n=1)$. Again we omit
arbitrary constant $z_0$. Equation \eqref{ODE3_KW_4} admits one
formal Laurent series in a neighborhood of the pole $z=0$. This
series is the following
\begin{equation}
\begin{gathered}
\label{ODE_KW_4_Laurent_As} w(z)=\frac{280}{ z^4}
-\frac{140\beta}{39 z^2}+\frac{507C_0-31\beta^2}{3042}
-\frac{31\beta^3z^2}{237276} +\ldots
\end{gathered}
\end{equation}
The Fuchs indices for this expansion are $-1$,
$(11\pm\sqrt{159}i)/2$. Consequently, there are no arbitrary
constants in \eqref{ODE_KW_4_Laurent_As}. General expression for an
elliptic solution (see \eqref{Ex_Sol_Elliptic2}) can be written as
\begin{equation}\begin{gathered}
\label{ODE_KW_4_Sol_el1}
w(z)=\frac{c_{-4}}{6}\,\wp_{zz}+c_{-2}\,\wp+h_0,\quad
c_{-4}=280,\quad c_{-2}=-\frac{140\beta}{39}.
\end{gathered}
\end{equation}
Necessary condition $c_{-1}=0$ is automatically satisfied. Finding
the Laurent series for this function in a neighborhood of the point
$z=0$, we get
\begin{equation}
\label{ODE_KW_4_Sol_el1LS} w(z)=
\frac{c_{{-4}}}{{z}^{4}}+\frac{c_{{-2}}}{{z}^{2}}+{\frac
{c_{{-4}}g_{{2}}}{60}}+h_ {{0}}+ \left(
\frac{c_{{-4}}g_{{3}}}{14}+\frac{c_{{-2}}g_{{2}}}{20} \right) {z}^
{2}+o \left( |z|^2\right),\, 0<|z|<\varepsilon_1.
\end{equation}
Comparing coefficients of this series with coefficients of expansion
\eqref{ODE_KW_4_Laurent_As}, we find the parameters $h_0$, $g_3$
\begin{equation}\begin{gathered}
\label{ODE_KW_4_Sol_el1P} h_0={\frac {C_{{0}}}{6}}-{\frac
{14\,g_{{2}}}{3}}-{\frac {{ 31\beta}^{2}}{3042}},\quad g_3={\frac
{7\beta\,g_{{2}}}{780}}-{\frac {31{\beta}^{3}}{4745520}}
\end{gathered}
\end{equation}
and the correlation on the parameters of equation \eqref{ODE3_KW_4}
\begin{equation}\begin{gathered}
\label{ODE_KW_4_Sol_el1C} C_2= \left({\frac {287}{207}}\,C_0^2+\frac
{1148}{69}\,C_1 -{\frac {41328}{656903}}\beta^{4}\right) g_{
{2}}+{\frac {372}{4826809}}\, \beta^{6} \\
\\
+{\frac
{5}{18252}}\,C_0^{2}\beta^{2}+{\frac {5}{1521}}\,\beta^2C_1-{\frac
{{C_{{0}}}^{3}}{108}}- {\frac {C_{{0}}C_{{1}}}{6}}.
\end{gathered}
\end{equation}
In expressions \eqref{ODE_KW_4_Sol_el1P}, \eqref{ODE_KW_4_Sol_el1C}
$g_2$ is a solutions of quadratic equation
\begin{equation}\begin{gathered}
\label{ODE_KW_4_Sol_el1g2}g_2^2 -{\frac
{{\beta}^{2}}{1014}}\,g_2+{\frac {1457{\beta}^{4}}{662158224}}-
{\frac {{C_{{0}}}^{2}}{23184}}-{\frac {C_{{1}}}{1932}}=0.
\end{gathered}
\end{equation}
Simplifying \eqref{ODE_KW_4_Sol_el1}, we obtain elliptic solutions
of equation \eqref{ODE3_KW_4}
\begin{equation}\begin{gathered}
\label{ODE_KW_4_Sol_el1e} w(z)=280\wp^{2} -{\frac {140\beta
}{39}}\wp+{\frac {C_{{0}}}{6}}-{\frac
{31}{3042}}\,\beta^2-28\delta\,g_2.
\end{gathered}
\end{equation}
Substituting this expression into  equation \eqref{ODE3_KW_4}, we
see that equation \eqref{ODE3_KW_4} indeed possesses solutions of
the form \eqref{ODE_KW_4_Sol_el1e} provided that correlations
\eqref{ODE_KW_4_Sol_el1C}, \eqref{ODE_KW_4_Sol_el1g2} hold.

Now let us construct simply periodic solutions. Expression
\eqref{Ex_Sol_Expp} with $p=4$ yields
\begin{equation}\begin{gathered}
\label{ODE_KW_4_Sol_per1}
w(z)=-\sqrt{L}\left\{\frac{c_{-4}}{6}\frac{d^3}{dz^3}-c_{-2}\frac{d}{dz}\right\}
\cot(\sqrt{L}z)+h_0,\quad \sqrt{L}=\frac{\pi}{T}.
\end{gathered}
\end{equation}
Expanding this function around the point $z=0$, we obtain
\begin{equation}\begin{gathered}
\label{ODE_KW_4_Sol_per1LS} w(z)=
\frac{c_{{-4}}}{{z}^{4}}+\frac{c_{{-2}}}{{z}^{2}}+\frac{c_{-4}L^2}{45}+\frac{c_{-2}L}{3}
+h_0+\frac{(20c_{-4}L+63c_{-2})L^2}{945}\, {z}^ {2}\\
\\
+o \left( |z|^2\right),\quad 0<|z|<\varepsilon_2.
\end{gathered}\end{equation}
Comparing coefficients of the series \eqref{ODE_KW_4_Sol_per1LS} and
\eqref{ODE_KW_4_Laurent_As}, we find the parameters $L$, $h_0$ and
correlations on the parameters of equation \eqref{ODE3_KW_4}.
Verification shows that equation \eqref{ODE3_KW_4} possesses
solutions of the form
\begin{equation}\begin{gathered}
\label{ODE_KW_4_Sol_per1e} w(z)=280{L_j}^{2}\cot^4 \left( \sqrt
{L_j}z \right)+ \frac{(140 L_j-\beta)L_j}{39}\cot^2 \left( \sqrt
{L_j}z \right) + {\frac {784{L_j}^{2}}{9}}\\
\\
+{\frac {C_{{0}}}{6}}-{\frac {31{\beta}^{2}}{3042}}-{\frac {280
L_j\beta}{117}},\quad L_j=\frac{\beta\kappa_j}{52},\quad j=1,2,3.
\end{gathered}
\end{equation}
if the following correlations are valid
\begin{equation}\begin{gathered}
\label{ODE_KW_4_Sol_per1_C}C_1=\left({\frac
{38801}{34273200}}\,{\kappa}^{2}-{\frac
{4991}{3427320}}\,\kappa +{\frac {40889}{34273200}}\right)\beta^4-{\frac {{C_{{0}}}^{2}}{12}},\hfill\\
\\
C_2=\left({\frac
{11113903}{193072360000}}\,-{\frac
{951419}{579217080000}}\,\kappa^2-{\frac
{1944971}{57921708000}} \,\kappa\right)\beta^6\\
\\
+\left({\frac {4991}{20563920}}\,\kappa-{\frac
{38801}{205639200}}\,\kappa^2-{\frac
{40889}{205639200}}\,\right)C_0\beta^4+{\frac {{C_
{{0}}}^{3}}{216}}.
\end{gathered}
\end{equation}
Three values for the parameter $\kappa$ are the following
\begin{equation}\begin{gathered}
\label{ODE_KW_4_Sol_per1_P} \kappa_1=-1,\quad
\kappa_2=\frac{31+3\sqrt{31}i}{20},\quad
\kappa_3=\frac{31-3\sqrt{31}i}{20}.
\end{gathered}
\end{equation}
Now let us construct rational solutions of equation
\eqref{ODE3_KW_4}. Substituting $w=z^m$ as $z$ tends to infinity
into equation \eqref{ODE3_KW_4}, we see that $m=0$. Note that at
certain conditions on the parameters $\beta$, $C_0$, $C_1$, $C_2$
expansion \eqref{ODE_KW_4_Laurent_As} terminates. As a result we
find a rational solution
\begin{equation}\begin{gathered}
\label{KW_4_Sol_rat} w(z)=\frac{180}{z^4}+\frac{C_0}{6},\quad
\beta=0,\quad C_1=-\frac{C_0^2}{12},\quad C_2=\frac{C_0^3}{216}.
\end{gathered}
\end{equation}
Elliptic and simply periodic solutions \eqref{ODE_KW_4_Sol_el1e}
\eqref{ODE_KW_4_Sol_per1e} were given in
\cite{Kudryashov10,Parkes01,Parkes02,Kano01}. Note that in the case
$\beta=0$ equation \eqref{ODE3_KW_4} does not have simply periodic
solutions, while at certain conditions on the parameters $C_0$,
$C_1$, $C_0$ this equation with $\beta=0$ possesses elliptic and
rational solutions.

\section{Exact solutions of equation \eqref{KW_4} with $n=2$}

In this section we construct meromorphic traveling wave solutions of
the modified Kawahara equation (equation \eqref{KW_4} with $n=2$).
Again we omit arbitrary constant $z_0$. The equation
\begin{equation}
\label{ODE3_KW_2}w_{zzz}w_z-\frac{1}{2}w_{zz}^2-\frac{\beta}{2}
w_z^2-30w^4+\frac{C_0}{2}w^2+C_1w+C_2=0.
\end{equation}
possesses two different formal Laurent series in a neighborhood the
pole $z=0$
\begin{equation}
\begin{gathered}
\label{ODE_KW_2_Laurent_As} w^{(1)}(z)=\frac{1}{
z^2}-\frac{\beta}{60}+\frac{10C_0-\beta^2}{3600}\,z^2+\frac{5C_0\beta+450C_1-\beta^3}{151200}\,z^4+
\ldots,\hfill \\
\\
w^{(2)}(z)=-\frac{1}{z^2}+\frac{\beta}{60}+\frac{\beta^2-10
C_0}{3600}\,z^2+\frac{\beta^3 +450C_1-5C_0\beta}{151200}\,z^4+
\ldots.\hfill
\end{gathered}
\end{equation}
The Fuchs indices for these expansions are the following $-1$,
$(7\pm \sqrt{71}i)/2$. Since none of them is a non--negative
integer, we see that all the coefficients in expansions
\eqref{ODE_KW_2_Laurent_As} are uniquely determined. First of all
let us construct exact meromorphic solutions having poles of single
type ($i=1$ or $i=2$). Note that equation \eqref{ODE3_KW_2}
possesses the symmetry
\begin{equation}
\label{ODE_KW_2_Sol_Symmetry}w(z;-C_1)=-w(z;C_1)
\end{equation}
Thus, without loss of generality, we need to find meromorphic
solutions with poles of the first type. Necessary condition for
existence of elliptic solutions is automatically satisfied. It
follows from the theorem 2 that we should take an elliptic
solution in the form
\begin{equation}
\label{ODE_KW_2_Sol_el1} w(z)=c_{-2}^{(1)}\wp(z;g_2,g_3)+h_0,\quad
c_{-2}^{(1)}=1.
\end{equation}
The Laurent series for this function in a neighborhood of the point
$z=0$ is the following
\begin{equation}
\label{ODE_KW_2_Sol_el1LS} w(z)= {\frac
{c_{{-2}}^{(1)}}{{z}^{2}}}+h_{{0}}+\frac{c_{{-2}}^{(1)}g_{{2}}{z}^{2}}{20}\,
+\frac{c_{{-2}}^{(1)}g_{{3}}{z}^{4}}{28} \,+o(|z|^4),\quad
0<|z|<\tilde{\varepsilon}_1.
\end{equation}
Comparing coefficients of this series with coefficients of the
series $w^{(1)}(z)$ (see \eqref{ODE_KW_2_Laurent_As}), we find the
parameters of elliptic solution \eqref{ODE_KW_2_Sol_el1}
\begin{equation}
\begin{gathered}
\label{ODE_KW_2_Sol_el1_P} h_0=-\frac{\beta}{60},\quad g_2={\frac
{10\,C_{{0}}-{\beta}^{2}}{180}},\quad g_3={\frac {5\,C_{{0}}\beta\,+
450\,C_{{1}} -{\beta}^{3}}{5400}}
\end{gathered}
\end{equation}
and the correlation on the parameters of equation \eqref{ODE3_KW_2}
\begin{equation}
\begin{gathered}
\label{ODE_KW_2_Sol_el1_C} C_2={\frac
{32\,{\beta}^{4}-220\,C_{{0}}{\beta}^{2}
+125\,{C_{{0}}}^{2}-8100\,C_{{1}}\beta\,}{324000}}.
\end{gathered}
\end{equation}
Substituting obtained elliptic solution and the correlation
\eqref{ODE_KW_2_Sol_el1_C} into equation \eqref{ODE3_KW_2}, we see
that it indeed satisfies the equation provided that
\eqref{ODE_KW_2_Sol_el1_C} holds. The explicit expression for simply
periodic solution with the first type of poles is the following
\begin{equation}\begin{gathered}
\label{ODE_KW_2_Sol_per1} w(z)=-\sqrt{L}c_{-2}^{(1)}\frac{d}{d\,z}
\cot(\sqrt{L}z)+h_0 \equiv Lc_{-2}^{(1)}
[\cot^2(\sqrt{L}z)+1]+h_0,\\
\\
\sqrt{L}=\frac{\pi}{T},\quad c_{-2}^{(1)}=1.
\end{gathered}
\end{equation}
The Laurent series for this function in a neighborhood of the point
$z=0$ can be written as
\begin{equation}
\label{ODE_KW_2_Sol_per1LS} w(z)= {\frac
{c_{{-2}}^{(1)}}{{z}^{2}}}+\frac{Lc_{{-2}}^{(1)}}{3}+h_{{0}}+\frac{{L}^{2}c_{{-2
}}^{(1)}\,{z}^{2}}{15}+{\frac
{2{L}^{3}c_{{-2}}^{(1)}\,{z}^{4}}{189}}\, +o(|z|^4),\quad
0<|z|<\tilde{\varepsilon}_2.
\end{equation}
Comparing coefficients of this series with coefficients of the
series $w^{(1)}(z)$ (see \eqref{ODE_KW_2_Laurent_As}), we determine
the parameters of simply periodic solution \eqref{ODE_KW_2_Sol_per1}
\begin{equation}
\begin{gathered}
\label{ODE_KW_2_Sol_per1_P} L=\frac{\kappa_j}{60},\quad
h_0=-\frac{\kappa_j+3\beta}{180},\quad  \kappa_j=\pm
\sqrt{15(10C_0-\beta^2)},\quad j=1,2
\end{gathered}
\end{equation}
and correlations on the parameters of equation \eqref{ODE3_KW_2}
\begin{equation}
\begin{gathered}
\label{ODE_KW_2_Sol_per1_C} C_1={\frac
{10\,\kappa_j\,C_{{0}}-\kappa_j\,{\beta}^{2}-45\,\beta\,C_{{0}}
+9\,{\beta}^{3}}{4050}}, \hfill \\
\\
C_2={\frac
{14\,{\beta}^{4}-130\,{\beta}^{2}C_{{0}}+125\,C_{{0}}^2+2\,{\beta}^{3}\kappa_j-20\,
\beta\,\kappa_j\,C_{{0}}}{324000}}.
\end{gathered}
\end{equation}
Checking--up obtained solutions, we see that equation
\eqref{ODE3_KW_2} indeed possesses solutions of the form
\begin{equation}
\label{ODE_KW_2_Sol_per1_res} w(z)=\frac{\kappa_j}{60}
\cot^2\left(\frac{\sqrt{15\kappa_j}}{30}\,
z\right)+\frac{2\kappa_j-3\beta}{180},\quad j=1,2,
\end{equation}
if correlations \eqref{ODE_KW_2_Sol_per1_C} hold. In the same way we
obtain the rational solution with the pole of the first type. The
rational function
\begin{equation}
\label{ODE_KW_2_Sol_rat1} w(z)={\frac {1}{{z}^{2}}}-{\frac
{\beta}{60}}.
\end{equation}
solves equation \eqref{ODE3_KW_2} provided that the following
correlations hold
\begin{equation}
\label{ODE_KW_2_Sol_C} C_0=\frac{\beta^2}{10}, \quad
C_1=\frac{\beta^3}{900}, \quad C_2=\frac{\beta^4}{144000}.
\end{equation}
Meromorphic solutions with poles of the second type may be found
with the help of the symmetry \eqref{ODE_KW_2_Sol_Symmetry}. Now let
us construct meromorphic solutions that posses poles of two types at
the same type. Without loss of generality, let us suppose that the
point $z=0$ is a pole of the first type. The expression for elliptic
solutions is the following
\begin{equation}
\label{ODE_KW_2_Sol_el2}
w(z)=c_{-2}^{(1)}\wp(z;g_2,g_3)+c_{-2}^{(2)}\wp(z-a;g_2,g_3)+\tilde{h}_0,\,
c_{-2}^{(1)}=1,\, c_{-2}^{(2)}=-1.
\end{equation}
Expanding this function in a neighborhood of the points $z=0$, $z=a$
and introducing notation $A\stackrel{def}{=}\wp(A)$,
$B\stackrel{def}{=}\wp_z(a)$, yields
\begin{equation}\begin{gathered}
\label{ODE_KW_2_Sol_Sol_el2LS} w(z)={\frac
{c_{-2}^{(1)}}{{z}^{2}}}+c_{-2}^{(2)}A+\tilde{h}_{{0}}-c_{-2}^{(2)}Bz
+ \,o(|z|),\quad
0<|z|<\tilde{\varepsilon}_3,\hfill \\
\\
w(z)={\frac {c_{-2}^{(2)}}{ \left( z-a \right) ^{2}}} +
c_{-2}^{(1)}A+\tilde{h}_{{0}}+c_{-2}^{(1)}\left( z-a \right) + \,
o(|z-a|),\\
 0<|z-a|<\tilde{\varepsilon}_4.
\end{gathered}\end{equation}
Note that using an addition formula for the Weierstrass elliptic
function we can rewrite expression \eqref{ODE_KW_2_Sol_el2} as
\begin{equation} \label{ODE_KW_2_Sol_el2n}
w(z)=(c_{-2}^{(1)}-c_{-2}^{(2)})\,\wp +\frac{c_{-2}^{(2)}}{4}
\left(\frac{\wp_z+B}{\wp-A}\right)^2+\tilde{h}_0-c_{-2}^{(2)}A.
\end{equation}
Comparing coefficients of the series \eqref{ODE_KW_2_Sol_Sol_el2LS}
with coefficients of expansions $w^{(1)}(z)$, $w^{(2)}(z)$ (see
\eqref{ODE_KW_2_Laurent_As}), we obtain an algebraic system for the
parameters of meromorphic solution \eqref{ODE_KW_2_Sol_el2n}.
Solving this system added by equation $B^2=4A^3-g_2A-g_3$, we get
\begin{equation}
\begin{gathered}
\label{ODE_KW_2_Sol_el2_P} A=\frac{\beta}{60},\, B=0,\,
\tilde{h}_0=0,\, g_2={\frac {{\beta}^{2}+5\,C_{{0}}}{540}},\,
g_3=-{\frac {\beta\, \left( 2\,{\beta}^{2}+25\,C_{{0} } \right)
}{162000}}
\end{gathered}
\end{equation}
Equation \eqref{ODE3_KW_2} possesses an elliptic solution of the
form
\begin{equation}
\label{ODE_KW_2_Sol_el2nn}w(z)=\frac {10800\, \wp^2-360\,\beta\,\wp
+25 \,C_{{0}}-{\beta}^{2}}{180 \left( 60\, \wp-\beta \right)}
\end{equation}
provided that the following correlations hold
\begin{equation}
\begin{gathered}
\label{ODE_KW_2_Sol_el2_C}C_1=0,\quad C_2={\frac { \left(
11\,{\beta}^{2}+ 100\,C_{{0}} \right)  \left( 4\,{\beta}^{2} -25\,
C_{{0}}\right) }{1215000}}.
\end{gathered}
\end{equation}
According to theorem 2 we take an expression for simply
periodic solutions in the form
\begin{equation}
\label{ODE_KW_2_Sol_per2}
w(z)=Lc_{-2}^{(1)}[\cot^2(\sqrt{L}z)+1]+Lc_{-2}^{(2)}[\cot^2(\sqrt{L}(z-a))+1]+h_0,
\end{equation}
where $c_{-2}^{(1)}=1$, $c_{-2}^{(2)}=-1$, $\sqrt{L}=\pi/T$.
Expanding this function in a neighborhood of the points $z=0$, $z=a$
and introducing notation $A\stackrel{def}{=}\sqrt{L}\cot (
\sqrt{L}a)$, we obtain
\begin{equation}\begin{gathered}
\label{ODE_KW_2_Sol_per2LS} w(z)={\frac
{c_{-2}^{(1)}}{{z}^{2}}}+\frac{c_{-2}^{(1)}L}{3}+c_{-2}^{(2)}
\left(L +{A}^{2}\right)+h_{{0}}+2\,c_{-2}^{(2)}A \left( L+{A }^{2}
\right)z \hfill \\
 + \,o(|z|),\quad
0<|z|<\tilde{\varepsilon}_5,\\
\\
w(z)={\frac {c_{-2}^{(2)}}{ \left( z-a \right) ^{2}}}
+\frac{c_{-2}^{(2)}L}{3} +c_{-2}^{(1)}\left(L
+{A}^{2}\right)+h_{{0}}-2\,c_{-2}^{(1)}A \left( L+{A}^{2} \right)\\
\times\left( z-a \right) + \, o(|z-a|),\quad
0<|z-a|<\tilde{\varepsilon}_6.
\end{gathered}\end{equation}
Comparing coefficients of these series with coefficients of
expansions $w^{(1)}(z)$, $w^{(2)}(z)$ (see
\eqref{ODE_KW_2_Laurent_As}), we find the parameters of meromorphic
solution \eqref{ODE_KW_2_Sol_per2}
\begin{equation}
\label{ODE_KW_2_Sol_per2_P}
\begin{gathered}
L=\frac{\beta}{40},\quad A=0,\quad h_0=0
\end{gathered}
\end{equation}
and conditions on the parameters of equation \eqref{ODE3_KW_2}
\begin{equation}
\label{ODE_KW_2_Sol_per2_C}
\begin{gathered}
C_0=-\frac{11\beta^2}{100},\quad C_1=0,\quad C_2=0.
\end{gathered}
\end{equation}
In the case $A=0$ we get $a=\sqrt{10\beta}\pi /\beta$. Verification
shows that the following function
\begin{equation}
\label{ODE_KW_2_Sol_per2_res} w(z)=\frac{ \beta\,\cot
\left(\frac{\sqrt{10\beta}z}{10}\right)} {10\sin
\left(\frac{\sqrt{10\beta}z}{10}\right)}.
\end{equation}
indeed solves equation \eqref{ODE3_KW_2} with the parameters $C_0$,
$C_1$, $C_2$ given by \eqref{ODE_KW_2_Sol_per2_C}. If we try to find
rational solutions with two poles, we shall see that arising
algebraic system is inconsistent. Thus we have found the whole set
of non--constant meromorphic solutions for \eqref{ODE3_KW_2}. Note
that elliptic solutions \eqref{ODE_KW_2_Sol_el1},
\eqref{ODE_KW_2_Sol_el2nn} degenerate if the following condition is
valid $g_2^3-27g_3^2=0$.

Simply periodic solutions with poles of one type arise in
\cite{Kano01, Parkes01}. Elliptic solutions with poles of one type
were given in \cite{Parkes02}. Solutions with poles of different
types seem to be new.

\section{Exact solutions of equation \eqref{KW_4} with $n=4$}

In this section our goal is to find meromorphic traveling wave
solutions of the modified Kawahara equation (equation \eqref{KW_4}
with $n=4$). Recall that traveling wave solutions of equation
\eqref{KW_4} with $n=4$ and $\alpha$, $\delta$ given by
\eqref{Parameters} satisfy the following third order ordinary
differential equation
\begin{equation}
\label{ODE3_KW_1}w_{zzz}w_z-\frac12w_{zz}^2  -\frac{\beta}{2}
w_{z}^2-4w^6+\frac{C_0}{2}w^2+C_1w+C_2=0.
\end{equation}
This equation possesses the symmetries of the form
\begin{equation}
\label{ODE_KW_1_Sol_Symmetry}w(z;-C_1,C_2)=-w(z;C_1,C_2),\quad
w(z;iC_1,-C_2)=iw(z;C_1,C_2).
\end{equation}
Equation \eqref{ODE3_KW_1} admits four different formal Laurent
expansions in a neighborhood of the pole $z=0$. They are the
following
\begin{equation}
\begin{gathered}
\label{ODE_KW_1_Laurent_As} w^{(1)}(z)=\frac{1}{
z}-\frac{\beta}{60}z+\frac{15C_0-\beta^2}{180}z^3+ \frac{C_1}{96}z^4
\ldots,\hfill \\
\\
w^{(2)}(z)=-\frac{1}{
z}+\frac{\beta}{60}z+\frac{\beta^2-15C_0}{180}z^3+
\frac{C_1}{96}z^4+ \ldots.\hfill \\
\\
w^{(3)}(z)=\frac{i}{
z}-\frac{i\beta}{60}z+\frac{i(15C_0-\beta^2)}{180}z^3+
\frac{C_1}{96}z^4+
\ldots,\hfill \\
\\
w^{(4)}(z)=-\frac{i}{
z}+\frac{i\beta}{60}z+\frac{i(\beta^2-15C_0)}{180}z^3+
\frac{C_1}{96}z^4+ \ldots.\hfill
\end{gathered}
\end{equation}
All not written out coefficients are uniquely determined since the
Fuchs indices of these expansions are $-1$, $(5\pm \sqrt{39}i)/2$.
Let us construct exact meromorphic solutions with poles of one type.
Without loss of generality, we may consider solutions with poles of
the first type. Elliptic functions possessing one pole of the first
order in a parallelogram of periods do not exist. According to
theorem 2 we take an expression for simply periodic
solutions in the form
\begin{equation}
\label{ODE_KW_1_Sol_per1}
w(z)=\sqrt{L}c_{-1}^{(1)}\cot(\sqrt{L}z)+h_0,\quad
\sqrt{L}=\frac{\pi}{T},\quad c_{-1}^{(1)}=1.
\end{equation}
Following the procedure described in section \ref{Method applied}
and taking into account the symmetries
\eqref{ODE_KW_1_Sol_Symmetry}, we obtain
\begin{equation}
\label{ODE_KW_1_Sol_per1case1} w(z)=\pm \frac{\sqrt {5\beta}\cot
\left({\frac {\sqrt { 5\beta}}{10}z} \right)}{10 },\quad
C_0=\frac{3\beta^2}{50},\quad C_1=0,\quad C_2=\frac{\beta^3}{1000}
\end{equation}
and
\begin{equation}
\label{ODE_KW_1_Sol_per1case2} w(z)=\pm \frac{\sqrt {5\beta }i\cot
\left({\frac {\sqrt { 5\beta}}{10}z} \right)}{10 },\,
C_0=\frac{3\beta^2}{50},\, C_1=0,\, C_2=-\frac{\beta^3}{1000} .
\end{equation}
Again we omit arbitrary constant $z_0$. Rational solutions with one
pole can be written as
\begin{equation}
\label{ODE_KW_1_Sol_rat1} w(z)=\pm \frac{1}{ z},\quad w(z)=\pm
\frac{i }{ z},\quad \beta=0,\quad C_0=0,\quad C_1=0,\quad C_2=0.
\end{equation}
Now let us consider the case of meromorphic solutions with poles of
two different types. We begin with poles of the first and the second
type. It follows from theorem 2 that an elliptic solution
should be taken in the form
\begin{equation}
\label{ODE_KW_1_Sol_el1}
w(z)=c_{-1}^{(1)}\zeta(z;g_2,g_3)+c_{-1}^{(2)}\zeta(z-a;g_2,g_3)+\tilde{h}_0,\quad
c_{-1}^{(1)}=-c_{-1}^{(2)}=1.
\end{equation}
This function possesses two poles $z=0$, $z=a$ in a parallelogram of
periods. Expanding this function in a neighborhood of the points
$z=0$, $z=a$ and introducing notation $A\stackrel{def}{=}\wp(a)$,
$B\stackrel{def}{=}\wp_z(a)$,
$h_0\stackrel{def}{=}\tilde{h}_0-c_{-1}^{(2)}\zeta(a)$, we get
\begin{equation}\begin{gathered}
\label{ODE_KW_1_Sol_Sol_el1LS} w(z)={\frac
{c_{-1}^{(1)}}{{z}}}+h_0-c_{-1}^{(2)}Az+\frac{c_{-1}^{(2)}B}{2}z^2 +
\,o(|z|),\quad
0<|z|<\tilde{\tilde{\varepsilon}}_1,\hfill \\
\\
w(z)={\frac
{c_{-1}^{(2)}}{{z-a}}}+h_0-c_{-1}^{(1)}A(z-a)-\frac{c_{-1}^{(1)}B}{2}(z-a)^2
+ \,o(|z-a|),\\
0<|z-a|<\tilde{\tilde{\varepsilon}}_2.
\end{gathered}\end{equation}
In new variables elliptic solution \eqref{ODE_KW_1_Sol_el1} can be
rewritten as
\begin{equation}
\label{ODE_KW_1_Sol_el1n}
w(z)=\frac{c_{-1}^{(2)}(\wp_z+B)}{2(\wp-A)}+h_0.
\end{equation}
Comparing coefficients of the series \eqref{ODE_KW_1_Sol_Sol_el1LS}
with coefficients of expansions $w^{(1)}(z)$, $w^{(2)}(z)$ (see
\eqref{ODE_KW_1_Laurent_As}), we obtain the parameters of elliptic
solution \eqref{ODE_KW_1_Sol_el1n}
\begin{equation}
\begin{gathered} \label{ODE_KW_1_Sol_el1P}
A=-\frac{\beta}{60},\quad B=0,\quad h_0=0,\quad g_2={\frac
{\beta^{2}-10\,C_0}{120}},\\
g_3={\frac {\left(13\,{\beta}^{2}-150\,C_{ {0}} \right)\beta
}{108000}}.
\end{gathered}
\end{equation}
In addition several correlations on the parameters of equation
\eqref{ODE3_KW_1} arise. The elliptic solution and the correlations
on the parameters can be written as
\begin{equation}
\label{ODE_KW_1_Sol_el1nn}
w_{1,2}(z)=\frac{\wp_z}{2(\wp+\frac{\beta}{60})},\quad C_1=0,\quad
C_2={\frac {\left( 9\,{\beta}^{2}-100\,C_{{0} } \right)\beta
}{3000}}.
\end{equation}
Using symmetries \eqref{ODE_KW_1_Sol_Symmetry}, we obtain another
elliptic solution of equation \eqref{ODE3_KW_1}
\begin{equation}
\label{ODE_KW_1_Sol_el2nn}
w_{3,4}(z)=\frac{i\wp_z}{2(\wp+\frac{\beta}{60})},\quad C_1=0,\quad
C_2=-{\frac {\left( 9\,{\beta}^{2}-100\,C_{{0} } \right)\beta
}{3000}}.
\end{equation}
In the same way we find the simply periodic solution with poles of
the first and the second type. General expression for such a
solution (see theorem 2) reads as
\begin{equation}
\label{ODE_KW_1_Sol_per3}
w(z)=\sqrt{L}c_{-1}^{(1)}\cot(\sqrt{L}z)+\sqrt{L}c_{-1}^{(2)}\cot(\sqrt{L}(z-a))+h_0,\quad
\sqrt{L}=\frac{\pi}{T}.
\end{equation}
Equation \eqref{ODE3_KW_1} indeed possesses a solution of the form.
This solution is the following
\begin{equation}
\label{ODE_KW_1_Sol_per3e} w_{1,2}(z)= \frac{\sqrt{10\beta} } {10
\sinh \left(\frac{\sqrt{10 \beta}z}{10}\right)},\quad
C_0=\frac{9\beta^2}{100},\quad C_1=0,\quad C_2=0.
\end{equation}
Making use of symmetries \eqref{ODE_KW_1_Sol_Symmetry} we get the
simply periodic solution with poles of the third and the fourth type
\begin{equation}
\label{ODE_KW_1_Sol_per4e} w_{3,4}(z)=\frac{\sqrt{10\beta} i}
{10\sinh \left(\frac{\sqrt{10 \beta}z}{10}\right)},\quad
C_0=\frac{9\beta^2}{100},\quad C_1=0,\quad C_2=0.
\end{equation}
Algebraic systems for the parameters are inconsistent for all other
simply periodic solutions with poles of two different types as well
as simply periodic solutions with poles of three and four different
types. The same is true in the case of rational solutions with two,
three, and four poles. As far as elliptic solutions are concerned,
necessary condition \eqref{Condition_ellipticN} does not allow
existence of elliptic solutions with three poles in a parallelogram
of periods. In the case of elliptic solutions with four poles in a
parallelogram of periods the algebraic system for the parameters is
also inconsistent. Meromorphic solutions of equation
\eqref{ODE3_KW_1} obtained in this section seem to be new.

\section {Conclusion}

In this article we have studied traveling wave solutions of the
Kawahara and the modified Kawahara equations. We have found all the
families of meromorphic solutions for ordinary differential
equations \eqref{ODE3_KW_4}, \eqref{ODE3_KW_2}, \eqref{ODE3_KW_1}
describing the traveling wave solutions of the Kawahara equation and
its generalizations. In addition we have described a powerful method
for constructing exact meromorphic solutions (including elliptic,
simply periodic and rational solutions) of autonomous nonlinear
ordinary differential equations. Our method generalizes several
methods with an a priori fixed expression for an exact solution. Our
method allows one to find the whole set of exact meromorphic
solutions for a wide class of autonomous nonlinear ordinary
differential equations. Besides that our method may be useful if one
needs to classify meromorphic solutions of an autonomous nonlinear
ordinary differential equation. Indeed, our approach involves
calculation of the period and amount of poles in a stripe of periods
for simply periodic solutions and invariants $g_2$, $g_3$ of the
Weierstrass elliptic function $\wp(z)$ and amount of poles in a
parallelogram of periods for elliptic solutions.

\section {Acknowledgements}

This research was partially supported by Federal Target Programm
"Research and Scientific - Pedagogical Personnel of innovation in
Russian Federation on 2009 -- 2013.

%"Scientific and scientific-educational specialists of innovation
%Russia ФЦП "Научные и научно-педагогические кадры инновационной
%России"\, на 2009--2013 гг.

\end{document}